# Saturation of Stationary Inversion States in a Three-Level Traveling-Wave Quantum Amplifier with Bistable Resonator Pumping[*]

D.N. Makovetskii

*A. Usikov Institute of Radio Physics and Electronics, National Academy of Sciences of Ukraine
12, Academician Proskura St., Kharkov 61085, Ukraine*

The inversion states of a saturated three-level traveling-wave quantum paramagnetic amplifier have been investigated under conditions of bistable resonator pumping. The equations of motion for the vectorial order parameter have been obtained using adiabatic elimination of fast variables. These equations are a generalization of the scalar two-level Drummond model for the case of a three-level quantum active system. Isolated branches of the constant of inversion have been found from stationary solutions of equations for the vectorial order parameter. A analysis of bifurcations in the saturated quantum amplifier with bistable pumping has been represented with full details.

The saturation of transitions between quantum levels of active centers is of paramount importance at the excitation of states with inverse population difference of energy levels. This mechanism is the basis for operation of many devices of quantum electronics [1, 2] and other equipments of this type (for example, acoustical quantum amplifiers and phasers generators [3, 4], radio-frequency masers based on nuclear magnetic resonance [5] and other active systems [6]).

The investigation of nonlinear processes in optical quantum systems led to the experimental disclosure of effects of bistability and multistability [7] in the middle 70s. In the case of bistability there are two different stable states of a nonlinear optical resonator at the same set of control parameters. At the multistability there are more than two such stable states. The initial interest in optical bistable systems has been roused by the assumed possibility of their workability as a component basis for computers of a new generation [8], since the switching time between states of a bistable cell of the nonlinear optical resonator can be small (less than $10^{-9}$ s).

The microwave analogue of the optical bistability has been also observed in a number of cases in paramagnetic [9] and gas [7,10] operating media in the frequency range from 10 to 85 GHz, however the typical switching times for microwave bistable systems turn out to be much more orders greater than for optical ones. Therefore, to develop computer logic elements, the microwave bistable cells can not be used. Nevertheless the investigation of bistability in microwave systems is not only of scientific interest but sharply defined practical interest although from the absolutely different point of view.

The fact is that the ramification of states of a nonlinear system (the bistability and multistability of this phenomenon are particular cases) is the rule rather than the exception, if the corresponding nonlinear parameter $C$ exceeds the specific threshold. For the optical bistable system this threshold is well known $C_{opt} = 4$, where $C_{opt}$ is the parameter of cooperativity [7, 11] being proportional to the resonator figure of merit.

The similar threshold exists as well as for the microwave nonlinear resonator that was shown,

---

for example, for acoustic resonators with saturable paramagnetics[12]. It has been found experimentally [13, 14] that when in use a high-Q pumping resonator in the traveling-wave quantum paramagnetic amplifier (maser) the bistability appears on a frequency of pumping. The experiments outlined have been carried out using the amplifier with an active andalusite crystal at the signal frequency $\Omega_S = 43$ GHz and the pumping frequency $\Omega_P \approx 150$ GHz [13, 14]. The electrodynamic system in use ensured the traveling-wave mode on the frequency $\Omega_S$ and at the same time the standing-wave mode on the pumping frequency $\Omega_P$ (the signal and pumping fields had mutually orthogonal polarizations; at that the grid transmitting the signal field and reflecting the pumping field had been used) [13, 14]. As experiments shown [13, 14], at the same pumping power there are two stable stationary branches of the maser system containing a high-Q pumping resonator and a saturable paramagnetic. And so the inversion can be either large (the constant of inversion $K \approx K_M$, where $K_M$ is the maximally accessible value $K$) or small ($K << K_M$) depending on the prehistory of the system, that had been observed in the experiments [13, 14].

At that there is no way to get out of bistability without reducing the Q-factor of the pumping resonator. At the same time, to obtain $K \approx K_M$ at a lower Q-factor of the pumping resonator it is necessary to give a higher pumping power to the system than at the high Q-factor of this resonator. It can turn out to be absolutely unacceptable from the point of view of the normal operation of the cryogenic system (all quantum amplifiers operate at low temperatures [1, 2]; the pumping power capability is usually less than 1 W). In breaking the normal cooling of the crystal for inversion there is a need in more powerful pumping at a rise in temperature it is more difficultly to saturate the spin-system, since the time of longitudinal relaxation reduces). This concerns actually the pumping on the frequencies of the millimeter wave band, where the time of longitudinal relaxation is anyway quite small even at the liquid helium temperature.

Thus, the branching of stationary inversion states (particularly bistability) is an unavoidable concomitant of the nonlinear system containing a high-Q resonator with saturated paramagnetic. As a matter of fact, it is the price which we have to pay for possibility of reduction of the pumping power keeping the high constant of inversion. If the quantum amplifier works in normal conditions, that is, a weak signal comes to its input and intensive noise is absent, then the behavior of the pumping bistable system is entirely controllable. At the beginning of the experiment the necessary branch of inversion should be chosen and after that the amplifier works in the prescribed regime as long as it is necessary [13, 14].

However, under the real-life conditions there is always noise (including intensive one) which can also come to the input of the quantum amplifier resulting in more or less short-time saturation of the amplifier in the signal channel. Thus, for a certain time the three-level active system turns out to be saturated by the fields of two frequencies, at that, this saturation is nonadditive since two saturated quantum transitions has a common spin level. It is obvious that in this case the three-level system should be considered as a whole and not just the two-level pumping subsystem as it has been done in [13, 14]. In what way does the pumping system behave in this case? After all the state of the pumping system, and so the current value of $K$, depend on the prehistory of this system. Does the amplifier come back to the normal operating mode after stopping the saturating noise? What kind of pumping mode should be chosen for self-recovery of elevated inversion? The need for solution of these important practical questions has defined the problem definition of this paper.

The purpose of this work is to study the influence of saturation through the signal channel on the inversion states formed by the resonator pumping under conditions of its bistability. An analytical model of a class "B" active (inverse) system [6] is the subject of inquiry, where the times of longitudinal relaxation exceed essentially all other characteristic times of transient processes.

## 1. Matter for scientific inquiry

A simple model of the traveling-hypersound-wave quantum paramagnetic amplifier (the wavelength of hypersound is of order of 1 µm) with the bistable electromagnetic resonator pumping (in millimeter wave band) has been chosen as a concrete subject for investigation. This model consists of the Fabry-Perot quasi-optical electromagnetic resonator with sizes being much more than the pumping wave-length. The resonator is filled with dielectric monocrystal with paramagnetic centers (the number of spin levels is not less than three) and with high hypersound transparency. One of acoustical axes is directed along the surface of the mirrors of the Fabry-Perot electromagnetic resonator. The hypersound wave vector is directed, in turn, along the acoustical axis. In this case the traveling-wave mode of hypersound is provided through the signal channel and the appearance of bistability is possible in the pumping channel because of nonlinearity of paramagnetic susceptibility and inner back-coupling in the resonator at the pumping frequency [7].

A similar model describes the electromagnetic traveling-wave maser with delay of a signal wave at the expense of a high dielectric susceptibility of crystal. Lastly, under conditions of electromagnetic traveling-wave of the signal the resonator pumping can be also realized on the basis of other systems, for example, under conditions when polarizations of electromagnetic fields of the signal and pumping are perpendicular to each other and there is an electrodynamics system (for example, grating) permitting us to provide the reflection of the pumping field and the transmission of the signal field [13, 14].

In the system under consideration the saturation factor $Z(D,C,Y)$ on the spin transition of the pumping can be an ambiguous function of the normalized input amplitude $Y$ of the microwave pumping field. For this purpose the condition $C > 4$ should hold, where $C$ is the parameter of cooperativity [7, 11] having the form $C = Q_C^{(0)}/4Q_M = \xi/4$ [15 - 20] in our case. Here $Q_C^{(0)}$ is the loaded Q-factor of the pumping resonator out of the regions of the magnetic resonance; $Q_M$ is the magnetic Q-factor [1, 2] of the considered dissipative system (paramagnetic and microwave resonator) at the pumping frequency in the absence of saturation of the spin transitions; $\xi$ is the designation for the parameter of cooperativity in papers [16-20]. The value $D$ is the population difference normalized to its thermodynamic equlibrium value on the pumping transition.

We will suppose that the spin system of the active paramagnetic medium possesses three energy levels $E_1$, $E_2$, $E_3$ (the numeration is in ascending order of the energy), at that $E_3 - E_1 = \hbar\Omega_P$, $E_2 - E_1 = \hbar\Omega_S$. In the absence of detuning in the traveling-wave maser, where anyone high-Q resonators at the frequency of signal and frequency of no-load transition $\Omega_F = \Omega_P - \Omega_S$ are unavailable, we can restrict ourselves to the scalar adiabatic model of bistability.

This model based on the subjection of fast variables [21] has been suggested for the first time by Drummond for optical bistable systems [22]. The model mentioned has been studied in details for the case of microwave pumping systems of paramagnetic masers and phasers in the absence of spin-transition saturation of the signal [16-20]. Within the frame work of this model it is necessary that all longitudinal relaxation times $T_1^{(k)}$, $k \in \{P,S,F\}$ (that is on the pumping, signal, and free-running transitions) are much more than both the transverse relaxation times $T_2^{(k)}$ (usually in the active medium of the traveling-wave maser all transverse relaxation times are the values of one order: $T_2^{(k)} \approx T_2 \approx 10^{-7}$–$10^{-8}$ s) and the photon life time $T_{CAV}$ in the pumping resonator (there are no other resonators in this model). Under such conditions the intensive external disturbance with the frequency $\Omega_S$ can be taken into account in the form of some renormalizations for the variables of state of this system. We consider such case in details below.

## 2. Unsaturated traveling-wave maser

To be specific, we restrict ourselves to the case of the phonon traveling-wave maser (phaser hypersound amplifier) with a quasi-optical resonator pumping as in [19]. However the most of the results obtained below are valid for electromagnetic signal traveling-wave masers with resonator pumping as well. From now on, we use the modification of a bistable pumping model which has been suggested before in [12] and studied in papers [16-20] for the case of an unsaturated traveling-wave maser. The initial model [16-20] is semiclassical, as the most analogous models of optical bistability [7]. It is based on the Bonifacho-Lugiato approximation for the splitting variables of state [11], where it is assumed that the condition of $Q_C^{(0)}/Q_M \to \text{const} = \xi$ is fulfilled when

$$(1/Q_C^{(0)} \to 0) \wedge (1/Q_M \to 0).$$

In this approximation $D$ (the population difference on the saturated spin pumping transition $E_1 \leftrightarrow E_3$, normalized to its thermodynamic equlibrium value) is the parameter of the order. The inversion coefficient $K$ on the unsaturated signal transition $E_1 \leftrightarrow E_2$ is determined in accordance with the current value of $D$ (see formulae (6) and (11) in paper [16]). The model [16-20] is the summarizing of the two-level adiabatic Drummond model (see formulae (5) in paper [22]) for the case of a three-level system with detuning on the pumping frequency and magnetic field, and with the second unsaturating field. If the detuning mentioned above is lacking and in our three-level system the signal field remains unsaturating as before, formulae (6) and (11) from paper [16] can be put in the following simple form:

$$T_1^{(P)} \frac{dD}{dt} = 1 - D - \frac{Y^2 D}{(1+2CD)^2} ; \tag{1}$$

$$K_i = (K_M + 1)[1 - D_i^{(st)}(C, Y)] - 1, \tag{2}$$

where $K_M = \sup_{Y \to \infty}(K(Y))$ is the maximal value of the inversion coefficient in the three-level active system, depending on the certain properties of the traveling-wave maser spin-system (for example, in the case of the approximate equation of the probabilities of the spin-lattice relaxation for all the spin transitions of the active medium [2]; $K_M = L - 1$, where $L = \Omega_P/2\Omega_S$); $K_i$ is the branches of the inversion coefficient corresponding to the stationary solution $D_i^{(st)}$ of equation (1) and index $i$ runs through the values $i = 1, 2, 3$ in the general case (the numeration is in ascending order of $K$). However the physically realizable branches $D_i^{(st)}$ and accordingly $K_i = K(D_i^{(st)})$ are only these ones for which the following inequalities are valid:

$$0 < \text{Re}\,[D_i^{(st)}(L, C, Y)] \leq 1;$$
$$-1 \leq \text{Re}\,[K(D_i^{(st)})] < K_M;$$
$$\text{Im}[D_i^{(st)}(L, C, Y)] = \text{Im}[K(D_i^{(st)})] = 0.$$

The values $L$, $C$, $Y$ are the control system parameters in the space $\text{R}_+^3$, where $\text{R}_+$ is the set of nonnegative real numbers. To realize inversion it is necessary that the inequality $L > 1$ is fulfilled and the parameters $C$ and $Y$ can possess any nonnegative values bounded above by the characteristic values $C_{\max}$ and $Y_{\max}$ for each certain traveling-wave masers (in particular $Y_{\max} << T_1^{(P)}/T_2$). Recall that, by definition, $K = -D_S$ [2], where $D_S$ is the population difference on the signal spin transition $E_1 \leftrightarrow E_3$, normalized to its thermodynamic equilibrium value. At the effective pumping the almost complete saturation of the pumping transition $E_1 \leftrightarrow E_3$ is achieved, i.e. $D = \varepsilon$, where $\varepsilon \ll 1$. Hence, on the signal transition $E_1 \leftrightarrow E_2$ the value of the inversion coefficient which is close to the maximal one $K = -D_S \approx K_M$, is realized. In the case when pumping is lacking $D_S = 1$, i.e. $K = -1$. Thus the dynamics of the three-level traveling-wave maser with an unsaturating signal and bistable pumping assume the minimal specification in the space $M_{0,d} \otimes M_{0,c} \subset R_+^1 \otimes R_+^3$ (index "0" points out the absence of saturation through the signal channel). Here $M_{0,d}$ is the space of dynamic variables of the traveling-wave maser, which in this case is one-dimensional one, $M_{0,d} = \{D\}$; $M_{0,c}$ is the control parameter space of the traveling-wave maser $M_{0,c} = \{L, C, Y\}$. The extended expressions for $C$ and $Y$ have the following form (see [18], where the formulae for the values $\xi \equiv 4C$ and $P \equiv Y^2$ are represented)

$$C = \pi \hbar^2 \Omega_P^2 \gamma_P^2 f_{CAV} N T_2 T_{CAV} / 2(2s+1) k_B \theta \ ;$$
$$Y = \gamma_P H_W \sqrt{T_1^{(P)} T_2 T_{CAV} / 2 T_R} \ .$$

Here $\gamma_P$ is the effective gyromagnetic ratio on the transition $E_1 \leftrightarrow E_3$; $f_{CAV}$ is the fill factor of the pumping resonator; $N$ is the concentration of the active centers; $s$ is the spin (for our three-level system $s = 1$); $k_B$ is the Boltzmann constant; $\theta$ is the temperature of the thermostat; $H_W$ is the amplitude of the field magnetic component $\Omega_P$ at the input of the pumping resonator; $T_R$ is the delay time of the electromagnetic field on the length of the pumping resonator. The considered model of the traveling-wave maser is valid when the following ratios are fulfilled [18]:

$$T_R \ll T_{CAV} \ ;$$
$$\hbar \Omega_P \ll (2s+1) k_B \theta \ ;$$
$$\Omega_P T_R \ll Q_{CAV}^{(0)} / 4C \ ;$$
$$H_W \ll (\gamma_P T_2^{(P)})^{-1} \sqrt{2 T_R / T_{CAV}} \ ;$$
$$T_1^{(P)} \gg T_2, T_{CAV}, T_R .$$

These inequalities bound definitely the region of this model validity of the traveling-wave maser (in paper [22] only the inequality $T_1^{(P)} \gg T_2, T_{CAV}, T_R$ has been represented).

In this way in the absence of signal transition saturation the stationary inversion states $K = K_i$ are described as the linear transformation of the roots of the Drummond equation

$$1 - D^{(st)} - \frac{Y^2 D^{(st)}}{(1 + 2C D^{(st)})^2} = 0 \ . \tag{3}$$

Using the standard methods of the catastrophe theory [23], from (3) we find that at the simultaneous achieving both the critical values for the cooperativity parameter $C = C_0^{[2]} = 4$ and the pumping parameter $Y = Y_0^{[2]} = 3\sqrt{3}$, the co-dimension bifurcation 2 takes place in our system (the upper index in the square brackets points out the co-dimension bifurcation, the inferior index "0" indicates the absence of saturation of the three-level system on the signal channel). At $C > C_0^{[2]}$ the co-dimension bifurcations 1 are realized at the points $Y_{0,\uparrow}^{[1]}$ and $Y_{0,\downarrow}^{[1]}$ (where $Y_{0,\uparrow}^{[1]} > Y_{0,\downarrow}^{[1]}$) corresponding to the jump-like transitions $K_1 \to K_3$ and $K_3 \to K_1$, that is from the smaller stationary value of the inversion coefficient to the larger one and vice versa.

Expression (2) can be also written in more obvious form

$$K_i = \frac{K_M Z_i^{(st)}(C,Y) - 1}{Z_i^{(st)}(C,Y) + 1}. \tag{4}$$

Hence it follows that for the lower (smooth) branch ($i = 1$, $Z^{(st)} = Z_1^{(st)}$) the considerable inversion can be obtained only at comparatively large $K_M$, while for the upper (hard) branch ($i = 3$, $Z^{(st)} = Z_3^{(st)} \gg Z_1^{(st)}$) the inversion is large even at $K_M \approx 1$. As it follows from equation (1), the middle branch with $i = 2$ is unstable in view of that $\lambda|_{D=D_2} > 0$, where $\lambda|_{D=D_2}$ is the Lyapunov index of system on the branch $K_2$ (see [16, 18]):

$$\lambda|_{D=D_2} = \frac{1 - D_2^{(st)}}{T_1 Y^2} \left( \frac{\partial D_2^{(st)}}{\partial (Y^2)} \right)^{-1}.$$

The possibility to decrease the input amplitude of the pumping field $Y$ to the value $Y_{0,\downarrow}^{[1]} < Y \ll Y_{0,\uparrow}^{[1]}$ after transferring (at $Y \geq Y_{0,\uparrow}^{[1]}$) from the lower inversion branch to the upper one ($K_1 \to K_3$) is the most essential advantage of the bistable pumping system. After that the traveling-wave maser operates in the stationary mode, where to keep the inversion with $K \approx K_M$, a smaller pumping power is required than in the case of common monostable pumping [16-20].

We note that the unidirectional transitions $K_1 \to K_3$ and $K_3 \to K_1$ of the considered bistable system are macroscopic (in contrast to quantum transitions between spin levels $E_1 \leftrightarrow E_3$, $E_1 \leftrightarrow E_2$, $E_2 \leftrightarrow E_3$). Recall that in the framework of model (1), (2) the value $K$ is determined exactly by the value $D$, not making an reverse impact on the latter, that is $K = K(D)$ but $D \neq D(K)$. In other words the population difference $D$ on the pumping transition $E_1 \leftrightarrow E_3$ is the only dynamic variable of the problem for the linear quantum amplifier, whereas the population difference $D_S \equiv -K$ on the signal transition $E_1 \leftrightarrow E_2$ is the linear transformation of the population difference $D$ on the pumping transition $E_1 \leftrightarrow E_3$.

## 3. Saturated traveling-wave maser

If a considerable saturation of $E_1 \leftrightarrow E_2$ transition takes place on the signal cannel, model (1), (2), certainly, becomes inadequate, i.e. the spin system gets considerable disturbance at the same time on two spin transitions with the frequencies $\Omega_P$ and $\Omega_S$; at that the spin level $E_1$ is common for these transitions (i.e. the disturbances are non-additive).
Hence, the task of determination of the values $D$ and $K$ must be self-consistent, at that the mentioned values start to depend not only on $L$, $C$ and $Y$ but on the normalized intensity $J$ of the saturated acoustic signal. Thus, in the general case the desired motion equations must have the following form:

$$\begin{cases} T_1^{(P)} \dfrac{dD}{dt} = f_D(D,K;J,L,C,Y); \\ T_1^{(S)} \dfrac{dK}{dt} = f_K(D,K;J,L,C,Y), \end{cases} \quad (5)$$

where the expression for $J$ can be written in the form [19]

$$J = T_1^{(S)} T_2 [\omega_{1U}^{(S)}/2]^2 = \dfrac{I}{I_0}. \quad (6)$$

Here $\omega_{1U}^{(S)}$ is the acoustic analogue of the so-called Raby frequency [7], $\omega_{1U}^{(S)} = |\gamma_U U|$; $I$ is the microwave acoustic signal intensity; $I_0 = 2\rho' v_S^3 \hbar^2 / T_1^{(S)} T_2 |\Phi_S|^2$; $\gamma_U$ is the acoustic analogue of the common gyromagnetic ratio, $\gamma_U = k_U \Phi_S / \hbar$; $k_U = |\vec{k}_U| = \Omega_S / v_S$; $\vec{k}_U$ is the wave vector of hypersound; $U$ is the acoustic signal amplitude; $\Phi_S$ is the parameter of the spin-phonon bond at the signal frequency; $v_S$ is the hypersound speed, $\rho'$ is the crystal density. For the case of the trigonal crystal field and longitudinal hypersound propagating along the crystallographic triad axis $\vec{c}$ (the active medium of the ruby phaser [2-4]) in accordance with [24] at $\angle(\vec{H},\vec{c}) \neq 0$ the expression for $\Phi_S$ takes the form

$$\Phi_S = \dfrac{1}{2} \left| G_{33} (3\langle \psi_\beta | \hat{s}_z^2 | \psi_\alpha \rangle - s(s+1)\langle \psi_\beta | \psi_\alpha \rangle) \right|,$$

where $G_{33}$ is the tensor component of the spin-phonon interaction [24]; $\hat{s}_z$ is the projection of the spin operator $\hat{s}$ onto the axis $Oz$; $|\psi_\alpha\rangle$ is the wave function belonging to the spin level $E_\alpha$; $\langle\psi_\beta|$ is the wave function which is complex-conjugated with respect to the wave function $|\psi_\beta\rangle$ belonging to the spin level $E_\beta$.

For longitudinal ($l$), fast transverse ($tf$), and slow transverse ($ts$) hypersonic waves propagating along the crystallographic twofold axis $\vec{a}$ (the active medium of the phaser is in $\text{Ni}^{2+}:\text{Al}_2\text{O}_3$ [25]), as it follows from [24], at $\vec{H} \parallel \vec{c}$, i.e. at $|\psi_{\alpha,\beta}\rangle = |\pm 1\rangle$, the expressions for $\Phi_S$ take the form

$$\Phi_S^{(l)} = \dfrac{1}{4}(G_{11} - G_{12})^2 + G_{16}^2;$$
$$\Phi_S^{(ts)} = [G_{14} \sin\chi + (1/2)(G_{11} - G_{12})\cos\chi] + (G_{15}\sin\chi + G_{16}\cos\chi)^2;$$
$$\Phi_S^{(ts)} = [G_{14} \cos\chi - (1/2)(G_{11} - G_{12})\sin\chi] + (G_{15}\cos\chi - G_{16}\sin\chi)^2,$$

where $\chi$ is the angle between the polarization vector of transverse hypersound and the axis $\vec{a}$.

We emphasize that the value $K$ in (5) is the same dynamic variable like $D$, and $J$ is a new control parameter (for clearness all the control parameters in the right-hand parts of equations (5) are separated by semicolons). Hence the stationary branches of the vector (two-dimensional) parameter of the order of $\vec{\Delta} \stackrel{\text{def}}{=} (D,K)$ can undergo bifurcations at the sufficiently slow scan of the vector of the control parameters $\vec{\Theta} \stackrel{\text{def}}{=} (J,L,C,Y)$ in the four-dimensional space. At that, as a rule, the values $T_1^{(P)}$ and $T_1^{(S)}$ do not differ so much from each other that system (5) can be reduced to one equation. On the contrary, in a number of cases we can consider $T_1^{(P)} \approx T_1^{(S)} \approx T_1$ [2]. Thus, here we can not return to some scalar order parameter using the standard procedure of subjection of variables [21].

Equations (5) represent the minimal model of the traveling-wave maser with saturation through the signal channel. In the general case we should take into account the relaxation field of the spin system as well combined action in respect with the combined action of the fields $\Omega_P$ and $\Omega_S$ occurring at the frequency of free-running transition $\Omega_F = \Omega_P - \Omega_S$. However, in view of the fact that for the traveling-wave maser the inequalities $r_k \equiv T_2^{(k)}/T_1^{(k)} \ll 1$ are typical, where $k \in \{P,S,F\}$ (usually $r_k = 10^{-6}$–$10^{-8}$, see, for example, [1, 2, 20]) and the considered system does not possess a resonator at the frequency $\Omega_F = \Omega_P - \Omega_S$, for all that we can restrict ourselves to the two-dimensional parameter of the order of $\vec{\Delta} = (D,K)$. For that we should take into consideration the maximal acceptable values of parameters $Y$ and $J$ for this model of a traveling-wave maser, namely $r_P^{1/2} Y \ll 1$, $r_S J \ll 1$, that is usually executed in real traveling-wave masers with much more reserve.

The similar in appearance situation (although in the different context) takes place for a three-level laser generator as well [26], where in the absence of off-tunings the minimal model of generation also is based on some two-dimensional parameter of the order of $\vec{\delta}$. Nevertheless there is a fundamental difference between our system and laser system presented in paper [26]. For the traveling-wave maser with saturation through the signal channel the condition $T_1 \gg T_{CAV}, T_2$ is fulfilled (class "B" non-equilibrium system [6]), whereas in [26] it has been considered the case when $T_{CAV} \gg T_1, T_2$, that corresponds to the class "A" non-equilibrium system [6]. In paper [26] the inverse values $\sigma = 1/T_{CAV}$, $\gamma_\parallel = 1/T_1$, $\gamma_\perp = 1/T_2$ have been used. In these designations the class "A" system is determined by the inequality $\sigma \ll \gamma_\parallel, \gamma_\perp$.

Therefore, in model [26] the amplitudes $\alpha$ and $\beta$ of in-resonator pumping and signal fields are main dynamic variables (certainly the laser signal appears in the resonator without even an external saturating field). In our model the population difference $D$ and $K \equiv -D_S$ on the spin pumping and signal transitions are main variables. Therefore in [26] the parameter of the order of $\vec{\delta} \stackrel{\text{def}}{=} (\alpha,\beta)$ relates to the field characteristics of the system, whereas our parameter of the order $\vec{\Delta}$ represents mainly the condition of the active medium (paramagnetic center system). In other words in paper [26] the fast relaxing population differences of the system adjust to the behavior of the pumping and signal fields, whereas in our system, on the contrary, the slowly relaxing population differences determine the global dynamics and the field variables adjust to them. The similar difference of the lead processes in the class "A" lasers and in the class "B" paramagnetic active systems has important consequences not only from the point of view of the adequate model creation of a saturating traveling-wave maser but for interpretation of the nature of nonlinear effects in other class "B" systems, for example, in a non-autonomous acoustic quantum generator [25].

## 4. Motion equation

We find the explicit form of the functions $f_D(\vec{\Delta},\vec{\Theta})$ and $f_K(\vec{\Delta},\vec{\Theta})$ at the mentioned restrictions and at the configuration of the traveling-wave maser being similar to the one described in [19], but at simultaneous saturation of the traveling-wave maser through the signal and pumping channels. Since, in contrast to paper [19], there is not any detuning (in pumping frequency, magnetic field and so on) in the considered model, the constitutive equations of our task are simplified in comparison with equations (2,a) from paper [19] and comes to the balanced motion equations (see, for example, [2]):

$$\frac{dn_i}{dt} = \sum_{m \neq i}[(n_m - n_i)F_{im} + n_m W_{mi} - n_i W_{im}], \qquad (7)$$

where $n_i \equiv N_i/N$; $N_i$ is the filling numbers of the spin levels $E_i$, at that $N_1 + N_2 + N_3 = N$; $W_{im} = W_{mi}\exp[(E_i - E_m)/k_B\theta]$ is the probabilities (in a unit of time) of spin interaction with heat phonons [24] determining the times of the longitudinal relaxation $T_1^{(k)}$ (where $k \in \{P,S,F\}$) for each of the spin transitions $E_i \leftrightarrow E_m$, at that $W_{im} \ll T_{CAV}^{-1}, T_2^{-1}$; $F_{im}$ is the probabilities (in a unit of time) of spin system interaction with the saturating fields ($F_{im} = F_{mi}$), in our case, taking the following form

$$F_S = F_{12} = F_{21} = J/2T_1^{(S)}; \qquad (8)$$
$$F_P = F_{13} = F_{31} = Z/2T_1^{(P)}; \qquad (9)$$
$$F_F = F_{23} = F_{32} = 0, \qquad (10)$$

at that the essential distinction between $F_S$ and $F_P$ takes place. The fact is that $J$ is the control parameter depending on the spin system (on the signal channel the traveling-wave mode of operation is realized). At the same time the value $Z$ depends not only on the control parameters but on the current state of the spin system. Indeed, owing to the feedback appearing in the pumping resonator, as it follows from equations (7) in paper [19], when the detuning is available and after elimination of transverse magnetization of the spin system, the abridged wave equation for the pumping field takes the following form

$$2T_{CAV}\frac{dX}{dt} = Y - X - 2CDX, \qquad (11)$$

where $X = \sqrt{Z}$ is the normalized amplitude in-resonator pumping fields [7,11] which does not depend evidently on $J$ (but it depends on $D(J)$, that is typical for the all class "B" active systems) to the Bonifacho-Lugiato approximation and in the absence of detuning. Later on, in view of the fact that $T_{CAV} \ll W_{im}^{-1} \approx T_1^{(S)}, T_1^{(P)}$, at the joint examination of equations sets (5) and (11) the right-hand member of equation (11) can be set equal to zero. Then

$$Z = Z(D) = \frac{Y^2}{(1+2CD)^2}. \qquad (12)$$

Thus at the adiabatically slow change of the control parameters the factor of the pumping transition saturation, which is proportional to the squared amplitude of the in-resonator field, "watches" the current state of the population differences $D(J,L,C,Y)$ determined, in turn by equations (7). Let us consider these equations in detail. Expressing $n_1$, $n_2$, $n_3$ in terms of $n_P \equiv n_3 - n_1$ and $n_S \equiv n_2 - n_1$ and taking into consideration that $n_1 + n_2 + n_3 = 1$ we bring equation (7) to the following form

$$\frac{dn_P}{dt} = -2n_P F_P - n_S F_S + R_P ; \qquad (13)$$

$$\frac{dn_S}{dt} = -2n_S F_S - n_P F_P + R_S , \qquad (14)$$

where

$$R_P = \frac{1}{3}(\rho_0 - \rho_P n_P - \rho_S n_S); \qquad (15)$$

$$R_S = \frac{1}{3}(\sigma_0 - \sigma_P n_P - \sigma_S n_S). \qquad (16)$$

The explicit expression for $\rho_j$, $\sigma_j$, where $j \in \{0,P,S\}$, is found in the following form:

$$\rho_0 = W_{12} - W_{21} + W_{23} - W_{32} + 2W_{13} - 2W_{31};$$
$$\rho_P = W_{12} + 2W_{13} + W_{23} - W_{21} + 2W_{32} + 4W_{31};$$
$$\rho_S = W_{12} + 2W_{13} - 2W_{23} + 2W_{21} - W_{32} - 2W_{31};$$
$$\sigma_0 = 2W_{12} - 2W_{21} + W_{13} - W_{31} - W_{23} + W_{32};$$
$$\sigma_P = 2W_{12} + W_{13} - W_{23} - 2W_{21} - 2W_{32} + 2W_{31};$$
$$\sigma_S = 2W_{12} + W_{13} + 2W_{23} + 4W_{21} + W_{32} - W_{31}.$$

The formulae (13) - (16) describe the saturation in the three-level traveling-wave maser for arbitrary $\eta_k = \hbar\Omega_k / k_B \theta$, where $k \in \{P,S,F\}$. To simplify the problem, we assume that

$$-n_k^{(0)} \equiv \eta_k/(2s+1) \equiv \hbar\Omega_k/(2s+1)k_B\theta << 1,$$

where $n_k^{(0)}$ are the population difference under the thermodynamic equilibrium conditions of the spin system $Y = J = 0$ (under the condition all $n_k^{(0)}$ are negative ones). Moreover, we restrict ourselves to the case when $W_{12} = W_{13} = W_{23} = W$ (this means that $T_1^{(k)} \approx T_1 \equiv (3W)^{-1}$, $k \in \{P,S,F\}$). Then $\rho_j$, $\sigma_j$ are brought to the simple expressions

$$\rho_0 = -3\eta_P W ; \quad \sigma_0 = -3\eta_S W ; \qquad (17)$$
$$\rho_P = 9W ; \quad \sigma_S = 9W ; \qquad (18)$$
$$\rho_P = -3\eta_F W ; \quad \sigma_S = 0 . \qquad (19)$$

Let us introduce the dimentionless time $\tau = t/T_1$. Then, using $D \equiv n_P/n_P^{(0)}$, $K \equiv -n_S/n_S^{(0)}$, we obtain the extension of equations (1) and (2) in the case of an intensive field of the signal saturating the spin transition $E_1 \leftrightarrow E_2$ in the traveling-wave mode

$$\frac{dD}{d\tau} = f_D(\vec{\Delta},\vec{\Theta}) = 1 - D - \frac{Y^2 D}{(1+2CD)^2} + \frac{JK}{4L} ; \qquad (20)$$

$$\frac{dK}{d\tau} = f_K(\vec{\Delta},\vec{\Theta}) = -1 - K - JK + \frac{LY^2 D}{(1+2CD)^2} , \qquad (21)$$

where $T_1^{(k)} \approx T_1$ and there is an obvious restriction on the maximal value of the saturating signal intensity $J \ll T_1/T_2$. Strictly speaking there is a restriction on $J$ and from below: $J > |n_F^{(0)}| \equiv \hbar\Omega_F/(2s+1)k_B\theta$, however, to our approximation $|n_F^{(0)}| \ll 1$, i.e. at the consideration of the saturation effects we can assume that $J_{\min} \approx 0$.

## 5. Stationary states

Equations (20) and (21) describe the dynamics of the traveling-wave maser with a saturating signal in the space $\mathrm{M}_{J,d} \otimes \mathrm{M}_{J,c} \subset \mathrm{R}^2 \otimes \mathrm{R}_+^4$, what is minimally necessary to the investigations of non-stationary (transition) processes in the active system, that is the dependences $\vec{\Delta} = \vec{\Delta}(t)$. Here $\mathrm{M}_{J,d} = \{D,K\}$; $\mathrm{M}_{J,c} = \{J,L,C,Y\}$; $\mathrm{R}$ is the set of real numbers. As regards the stationary states of the system (20), (21), they can be represented as the renormalized solutions of Drummond's equation (3) with the additional control parameter $J$ as shown below. We introduce the coefficients of renormalization $\kappa_C$ and $\kappa_Y$:

$$\kappa_C = 1 - \frac{J}{4(J+1)L};$$

$$\kappa_Y = 1 - \frac{J}{4(J+1)}.$$

Then, using the substitutions $\widetilde{C} = \kappa_C C$; $\widetilde{Y}^2 = \kappa_Y Y^2$; $\widetilde{D} = D/\kappa_C$, at $d/dt = 0$ we obtain from (20) and (21) Drummond's modified equation

$$1 - \widetilde{D} - \frac{\widetilde{Y}^2 \widetilde{D}}{(1+2\widetilde{C}\widetilde{D})^2} = 0 \qquad (22)$$

and the expression for the stationary values of the inversion coefficient in the traveling-wave maser with bistable pumping and saturation through the signal channel

$$K_i = \frac{\frac{K_M + 1}{\kappa_Y(J)}\left(1 - \widetilde{D}_i^{(st)}(\vec{\Theta})\right) - 1}{1 + J} . \qquad (23)$$

Formula (23) can be represented in the more obvious form

$$K_i = \frac{1}{1+J} \cdot \frac{\frac{K_M}{\kappa_Y(J)} \widetilde{Z}_i^{(st)}(\vec{\Theta}) - 1}{\widetilde{Z}_i^{(st)}(\vec{\Theta}) + 1} \qquad (24)$$

or

$$K_i = \frac{1}{1+J} \cdot \frac{K_M X_i^2(\vec{\Theta}) - 1}{\kappa_Y(J) X_i^2(\vec{\Theta}) + 1}. \qquad (25)$$

Here $\widetilde{D}_i^{(st)}$ is the solution to equation (22) and the values $\widetilde{Z}_i^{(st)}(\vec{\Theta}) \equiv [\widetilde{D}_i^{(st)}(\vec{\Theta})]^{-1} - 1$ are the stationary branches of the renormalized saturation factor of the pumping transition, which, obviously, can be represented as well as modified ratio (12)

$$\widetilde{Z}_i^{(st)}(\vec{\Theta}) = \frac{\kappa_Y Y^2}{[1 + 2\kappa_C C \widetilde{D}_i^{(st)}(\vec{\Theta})]^2}.$$

As it follows from (22) the bifurcations of co-dimension 2 in the subspace of the control parameters $\{C, Y\}$ have the form

$$C_J^{[2]} = 4\kappa_C^{-1}(J, L); \quad Y_J^{[2]} = 3^{3/2} k_Y^{-1/2}(J),$$

and the bifurcations of co-dimension 1 for $Y$ (at fulfilling $\{J, L, C\} = \text{const}$) take the following form

$$Y_{J,\uparrow\downarrow}^{[1]} = \frac{1}{\sqrt{2\kappa_Y(J)}} \left[ \mu_J^{(1)}(J, L) \pm \mu_J^{(2)}(J, L) \right]^{1/2},$$

where

$$\mu_J^{(1)} = \widetilde{C}^2(J, L) + 10\widetilde{C}(J, L) + 1 \,;$$
$$\mu_J^{(1)} = \sqrt{\widetilde{C}(J, L)[\widetilde{C}(J, L) - 4]^3} \,.$$

The explicit expressions for $\widetilde{D}_i^{st}(J, L, C, Y)$ or $\widetilde{Z}_i^{(st)}(J, L, C, Y)$ are very complicated in contrast to the explicit expressions given above for the bifurcation sets. It is more convenient to use inverse dependences such as the Bonifacho-Lugiato formula (see [11] and also [15-19]). In particular, we obtain all three stationary branches $\widetilde{Z}_i^{(st)}(Y) \equiv \kappa_Y X_i^2$ from (22) in the form of the single-valued inverse function $Y = \phi_Y(\widetilde{Z}^{(st)})$, where

$$\phi_Y(\widetilde{Z}^{(st)}) = \sqrt{\frac{\widetilde{Z}_i^{(st)}}{\kappa_Y}} \left( 1 + \frac{2\kappa_C C}{1 + \widetilde{Z}_i^{(st)}} \right),$$

from whence we find the formulae permitting us to draw the amplitude dependences of the stationary in-resonator pumping field $X(Y)\big|_{J,L,C}$ and $X(J)\big|_{L,C,Y}$ using the following inverse single-valued functions (the indices $st$ are omitted later on):

$$Y = X\left\{1 + \frac{2C\left[1 - \dfrac{J}{4(J+1)L}\right]}{1 + \left[1 - \dfrac{J}{4(J+1)}\right]X^2}\right\}, \quad (26)$$

and

$$J = \frac{4[(Y-X)(1+X^2) - 2CX]}{2\left(4 - \dfrac{1}{L}\right)CX - (Y-X)(4+3X^3)}. \quad (27)$$

## 6. Analysis of the dependences $K(J)$

Ratio (26) is the extension of the Bonifacho-Lugiato formula in the case of the three-level system with two saturating fields and formulae (25) and (27) give the set of equations for determining $K(J)|_{L,C,Y}$. In general case scanning the surface sections of the bifurcations by the hyperplane $\{L,C,Y\}$, give four different regions $O_p$ (where $p \in \{I, II, III, IV\}$) of the saturated traveling-wave maser behaviour, which are separated by three critical points $Y_q$, where $q \in \{I, II, III\}$. To be specific, the next case when $L = \text{const}$, $C = \text{const}$ is considered. Then at the different $Y$ the dependences $K(J)$ can vary in the following way:

- region I is the monostability region $O_I$ with a small, or even negative, inversion coefficient of the saturated signal transition $K \ll K_M$ (the only branch $K(J)|_{Y \in O_I}$ is a "precursor" of the lower branch $K_1(J)|_{Y \in O_{II}}$ see below). This region is located on the interval $0 < Y < Y_I$, where the critical point $Y_I$ corresponds to the case when $K_2 = K_3$ (the appearance of a hard branch of the inversion coefficient at $J = \varepsilon$);

- region II is the anhysteresis bistability region $O_{II}$ on the interval $Y_I < Y < Y_{II}$, where the critical point $Y_{II}$ corresponds to the case when $K_2 = K_1$ (to the collision of the separatrix with the smooth branch of the inversion coefficient). In the region $O_{II}$ there is the only bifurcation of co-dimension 1 of $J_\downarrow^{[1]}$-type but there is no bifurcation of $J_\uparrow^{[1]}$-type. In this region $\text{Im}\,K_{1,2,3} = 0$ at $0 < J < J_\downarrow^{(st)}$ and $\text{Im}\,K_1 = 0$, $\text{Im}\,K_{2,3} \neq 0$ at $J > J_\downarrow^{(st)}$. Here the transition $K_3 \to K_1$ which is realized at the collision of the upper stable branch $K(J)$ with the separatrix $K_2$, is permitted, whereas the inverse transition $K_1 \to K_3$ is forbidden. Correspondingly, in this region the branch $K_3(J)$ is isolated at the slow scanning of $J$ and at $J(t=0) < J_\downarrow^{[1]}$; $\partial J/\partial t > 0$ the inverted state $K_3$ is overturned at the point $J_\downarrow^{[1]}$ in an irreversible way. After this all the time the inverse scanning of $J$ keeps the traveling-wave maser on the low-effective branch $K_1 \ll K_M$. The hard effect on the system (shaking the control parameters [12, 19, 20]) can be the only way to overcome this trap. Thus it is the most dangerous case from the point of view of the normal functioning the traveling-wave maser;

- region III is the hysteresis bistability region $O_{III}$ on the interval $Y_{II} < Y < Y_{III}$, where the critical point $Y_{III}$ corresponds to collapse of the bistable state. In the region $O_{III}$ there is already both bifurcations of co-dimension 1, namely $J_\downarrow^{[1]}$ and $J_\uparrow^{[1]}$. At the slow scanning of $J$ and at $J(t=0) < J_\downarrow^{[1]}$;

$\partial J / \partial t > 0$ the inverted state $K_3$, as well as in the region $O_{II}$, is overturned at the point $J_\downarrow^{[1]}$. However in contrast to the region $O_{II}$ the inverse scanning of $J$ in the region $O_{III}$ allow us to come back from the low-effective branch $K_1$ to the operating branch $K_3$, for which $K_3 \approx K_M$ at $J \approx 0$;

- region IV is the region $O_{IV}$, where the inversion monostable mode of the saturated traveling-wave maser is realized as well as in the region $O_I$, but now the unique branch $K(J)|_{Y \in O_{IV}}$ is the continuation of the upper branch $K_3(J)|_{Y \in O_{III}}$ at the transition $Y$ over the point $Y_{III}$.

## 7. Output computation of *K*(*J*)

Fig. 1-4 shows the dependences of $K(J)$ obtained for the case when $C = 5$ and $L = 2$ in the bistability regions $O_{II}$ (Fig. 1-2) and $O_{III}$ (Fig. 3-4).

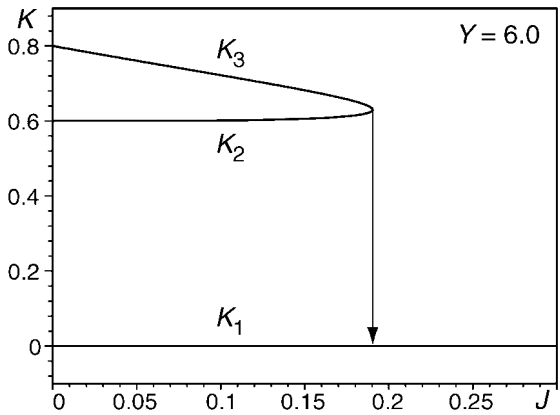 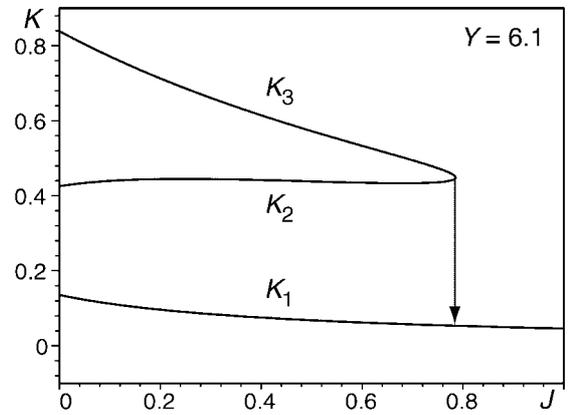

Fig. 1. The bistability of the inversion coefficient in the saturated traveling-wave maser at $L = 2$, $C = 5$, $Y = 6.0$ (the region $O_{II}$). The upper branch (with a high inversion coefficient) is isolated, the middle one is unstable, and the lower branch has practically zero-order inversion

Fig. 2. The bistability of the inversion coefficient in the saturated traveling-wave maser at $Y = 6.1$ (i.e. in the case when $Y = Y_{II} - \varepsilon$), when the approach to the hysteresis bistability region becomes visible. The upper branch has non-zero inversion

As follows from the figures the qualitative change in the behavior of $K(J)$, corresponding to the point $Y_{III}$, takes place on the interval $6.1 \leq Y \leq 6.2$. In Fig. 1-2 one can see clearly that the upper branch $K_3(J)$ is isolated, that is, at increasing $J$ it collides with the separatrix (the unstable middle branch $K_2$) and breaks at $J = J_\downarrow^{[1]}(Y)$, where $J_\downarrow^{[1]} \approx 0.2$ (Fig. 1) or $J_\downarrow^{[1]} \approx 0.8$ (Fig. 2). Owing to the image point of the system jumps on the lower branch $K_1$ (the inversion state is overturned by the saturating signal). There after the initial operation state (with the excited branch $K_3$) presents no possible to restore in the traveling-wave maser, if we change only the value $J$ and do not vary the other parameters, i.e. the image point of the system is situated on the branch $K_1$ all the time.

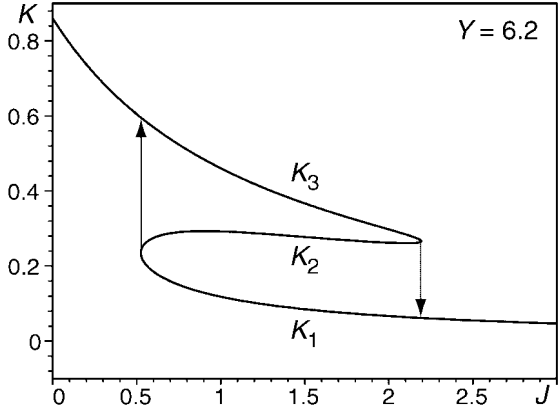 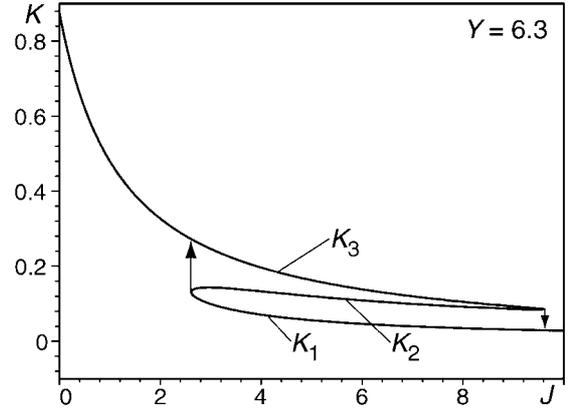

Fig. 3. The hysteresis bistability at $Y = 6.2$ (the region $o_{III}$). The parameters $L$ and $C$ are the same as in Fig. 1

Fig. 4. The hysteresis bistability in the region $o_{III}$ at $Y = 6.3$ (i.e. in the case when $Y = Y_{III} - \varepsilon$), when the hysteresis loop narrows sharply along the axis $K$ before outputting the system into the monostability region $o_{IV}$ (for which $Y > Y_{III}$). The parameters $L$ and $C$ are the same as in Fig. 1

As noted above now the inversion state can be thrown over on the upper branch $K_3$ only by means of hard excitation of the active system. For example, the exciting impulse should be given on the pumping channel [12, 19]. At that this impulse must be given every time after overturning the inversion state. Note that the analogous situation can emerge as well in the phaser generator with a high-Q pumping resonator [25], where instead of the effect of the external saturating acoustic field on the frequency $\Omega_S$, the self-impact of increasing induced phonon radiation takes place on this frequency.

Indeed the phonon field on the signal frequency (more specifically, on the eigen frequencies of the acoustical resonator of the phaser system), increasing at the self-excitation of such a generator, can lead to the overturn of the inversion from the upper branch $K_3$ to the lower branch $K_1$. After this, the generation can either break (at $K_1 < K_{tr} < K_3$, where $K_{tr}$ is the threshold value of the inversion coefficient at which the generation starts) or its power is very low due to the small ($0 < K_1 - K_{tr} << K_3 - K_{tr}$) self-excitation threshold crossing at $K_{tr} < K_1 < K_3$. At the pulse pumping in the system $Ni^{2+}:Al_2O_3$ the considerable (more than an order) increase of phaser generation intensity, which has been observed in paper [25], can be explained naturally on the basis of the model considered here. The pulse pumping restores the high value of $K \approx K_3$ at regular intervals, after the self-induced overturn of the inversion on the branch $K_1$, because of this the time-average intensity of the generated phonon flow increases considerably in comparison with the case of the unmodulated pumping of the phaser generator.

As a result of such hard periodic effect on the active system [12, 19], obviously, such considerable increase in the intensity of the generated phonon flow occurs, which we has discovered experimentally [25] in the phaser system $Ni^{2+}:Al_2O_3$ on fast transverse waves at $\Omega_S = 3$ GHz (the carrier pumping frequency $\Omega_P \approx 41.3$ GHz, the period of the pulse pumping modulation $T_m \leq 50$ μs). It should be added to this that the statement about the absence of generation in $Ni^{2+}:Al_2O_3$ at $K_{lin} > K_{tr}$ [27] (here $K_{lin}$ is the linear inversion coefficient), adduced by Peterson and Jacobsen before, can be also concerned with the effect of the inversion overturn on the branch $K_1$ when $K_1 < K_{tr} < K_{lin} \approx K_3$, considered in this paper.

At $C = 5$, $L = 2$ and $Y \geq 6.2$ (Fig. 3-4) the region, in which the bistability of the inversion coefficient is realized, is shifted to the direction of the greater values of $J$, at that the second critical point $J_\uparrow^{[1]} < J_\downarrow^{[1]}$ appears, corresponding to the collision of the branch $K_1$ with the separatrix ($J_\uparrow^{[1]} \approx 0.5$ in Fig. 3 and $J_\uparrow^{[1]} \approx 2.3$ in Fig. 4). In this case the bistability has a hysteresis nature what ensures the self-recovery of the initial mode of the traveling-wave maser after stopping the saturation through the signal channel ($J \to 0$).

## 8. Discussion of the results

The results under discussion show that in order for to avoid the "seizure" of the active system on a lower inversion branch and at the same time to operate at a minimally possible pumping level, the fine adjustment of the control parameters (first of all the input amplitude of the pumping field $Y$) is essential for a traveling-wave maser with the high-Q pumping resonator and external saturating signal. For the phaser generator in which the saturating field arises in the active crystal without an external signal, the proper pulse pumping mode can be chosen, ensuring the periodical recovery of the high inversion coefficient.

Using the above-mentioned system $Ni^{2+} : Al_2O_3$ [25] as an example, we estimate the values of the hypersound intensity $I$ at which the values of the parameter $J \approx 10$ are reached. Using ratio (6) and experimental data from paper [25] for the fast transversal hypersound propagating along the axis $\vec{a}$ at $J = 10$; $\rho' = 4$ g/cm$^3$; $v_S^{(tf)} = 6.7 \cdot 10^5$ cm/s; $\left|\Phi_S^{(tf)}\right|^2 = 270$ cm$^{-2}$ = $1.08 \cdot 10^{-29}$erg$^2$; $T_1^{(S)} = 0.02$ s; $T_2 = 10^{-8}$ s, we obtain $I \approx 1.2$ mW/cm$^2$.

Thus, the "overturn" of the inversion state on the branch with a small $K$ can occur even at comparatively low intensities of the saturating signal. Therefore, in our experiments [25] the intensity of the induced phonon radiation in the system $Ni^{2+} : Al_2O_3$ without modulation, is relatively low and, as it can be assumed, the impulse modulation ensures the periodical return onto the branch with a large $K \approx K_3$. As a result, the average power of the phaser generation (on transverse waves) in the system $Ni^{2+} : Al_2O_3$ increases by more than an order [25] at the pumping modulation, although in this case the average pumping power decreases significantly.

In the Bonifacho-Lugiato standard model (two-level system with one saturating field) in the absence of detuning of the active system, the isolated branching states are forbidden, as well-known from [7, 11], but the bictability is always hysteresis in nature. From this point of view, the behavior of the three-level system with two saturating fields is much more complicated even for the above-considered case of the balanced approximation and adiabatic exclusion of field variables. Note that the model of the three-level traveling-wave maser considered in this paper can be a basis for studying nonlinear processes in other active systems (excepting microwave and maser ones) [5, 6, 21, 28] in which there are three-level centers and mechanisms ensuring the appearance of one or another threshold phenomena as well. The obtained results correlate somewhat with conclusions of paper [29] in which from the quite different point of view the questions about the possibility of strong, irreversible, and mutual influence of the registered objects and measuring system are touched upon. Finally, it should be noted that the approach developed in this paper, can be summarized even for the more complicated case of two-resonator systems [25, 30-32], with the purpose of the further study of mechanisms of non-stationary and induced phonons in autonomous and non-autonomous phasers as well as in other class *"B"* multilevel active systems in which the longitudinal relaxation time of the active (excited) medium exceeds significantly the life times of electromagnetic and acoustical excitations.

## 9. Conclusions

For the model of the quantum amplifier with a signal traveling wave and high-Q (bistable) pumping resonator, it is shown that the saturation through the signal channel (for example, under the influence of noise) can lead to the step-wise transition of the active system on the branch with a low inversion coefficient. The analytic expressions describing this threshold effect have been obtained. The conditions have been found, under which the self-recovery of a high inversion occurs after stopping noise. The conditions, under which such a self-recovery does not take place, have been determined. The recommendations for choosing the quantum gain modes with the inversion self-recovery have been given. The possible mechanism of quantum generation with the periodical and forced inversion recovery has been considered, explaining the effect of increase in the intensity of phaser radiation discovered experimentally before in the system $Ni^{2+}:Al_2O_3$ [25].

The author thanks Makovetsky E.D. and Makovetskiy S.D. for the invaluable help in the work as well as Lavrov S.S. for giving a chance to become familiar with his article [29] before its publication, the results of which activate the fulfillment of these investigations.